\documentclass[%
 aip,
jcp,%
 amsmath,amssymb,
 reprint,%
]{revtex4-1}

\usepackage{graphicx}
\usepackage{dcolumn}
\usepackage{bm}
\usepackage{xcolor}

\newcommand{\Tr}{\mathrm{Tr}}
\newcommand{\tr}{\mathrm{tr}}
\newcommand{\rd}{\mathrm{d}}
\newcommand{\ii}{\mathrm{i}}
\newcommand{\eu}[1]{\mathrm{e}^{#1}}

\newcommand{\pd}[2]{\frac{\partial #1}{\partial #2}}
\newcommand{\thalf}{\ensuremath{\tfrac{1}{2}}}

\usepackage{braket}
\newcommand{\ketbra}[2]{\ket{#1}\!\bra{#2}}
\usepackage{dcolumn} 
\newcolumntype{d}[1]{D{.}{.}{#1}} 
\newcommand{\head}[1]{\multicolumn{1}{c}{#1}}

\begin{document}


\title{Spin-mapping approach for nonadiabatic molecular dynamics}

\author{Johan E. Runeson}
\email{johan.runeson@phys.chem.ethz.ch}
\author{Jeremy O. Richardson}%
\email{jeremy.richardson@phys.chem.ethz.ch}
\affiliation{Laboratory of Physical Chemistry, ETH Z\"{u}rich, 8093 Z\"{u}rich, Switzerland}

\date{\today}

\begin{abstract}
We propose a trajectory-based method for simulating nonadiabatic dynamics in molecular systems with two coupled electronic states.
Employing a quantum-mechanically exact mapping of the two-level problem to a spin-\thalf\ coherent state,
we construct a classical phase space of a 
spin vector constrained to a spherical surface
with a radius consistent with the quantum magnitude of the spin.
In contrast with the singly-excited harmonic oscillator basis used in Meyer-Miller-Stock-Thoss (MMST) mapping,
the theory requires no additional projection operators onto the space of physical states.
When treated under a quasiclassical approximation, we show that the resulting dynamics is equivalent to that generated by the MMST Hamiltonian. What differs is the value of the zero-point energy parameter as well as the initial distribution and the measurement operators.
For various spin-boson models the results of our method are seen to be a significant improvement compared to both standard Ehrenfest dynamics and linearized semiclassical MMST mapping,
without adding any computational complexity.


\end{abstract}

\maketitle

\section{\label{sec:level1}Introduction}
The simulation of electronically nonadiabatic dynamics is important for a wide range of applications across physics, chemistry and biology,
including the study of solar cells, vision, photosynthesis, radiation damage and many more.\cite{Tully2012perspective} What characterizes these nonadiabatic processes is that the electronic and nuclear motion are coupled, which causes the Born-Oppenheimer approximation to break down.
The computational task is particularly challenging in condensed-phase systems, where the number of nuclear degrees of freedom is typically too large for an exact quantum solution to be feasible. Approximate methods are therefore necessary to reach a reasonable balance between accuracy and computational cost.

This has led to the ongoing development of mixed quantum-classical dynamics,\cite{Stock2005nonadiabatic}
in which the nuclear degrees of freedom (called the environment)
are treated by classical molecular dynamics, and the evolution of the electronic states (called the subsystem) is treated quantum mechanically. 
It is, however, not trivial to treat the coupling between the two in a rigorous manner.
Popular methods such as Tully's fewest-switches surface hopping are computationally cheap but only heuristically motivated.\cite{Tully1990hopping,subotnik2013QCLE,martens2019,kapral2015quantum}
Another simple approach is to neglect all dynamical correlations between the classical environment and the quantum subsystem, which is called the \emph{Ehrenfest} or \emph{mean-field trajectory} method.\cite{ehrenfest1927,mclachlan1964} 

To go beyond Ehrenfest dynamics, a number of methods are based on mapping the quantum-mechanical system to an equivalent problem, for which one can find a well-defined classical limit. In this way, the dynamics of the quantum and classical degrees of freedom can be treated on an equal footing. Among the most commonly used examples is the Meyer-Miller-Stock-Thoss (MMST) mapping,\cite{Meyer1979nonadiabatic,Stock1997mapping} which maps the $N$ electronic levels to an $N$-dimensional singly-excited harmonic oscillator,
inspired by Schwinger's theory of angular momentum.\cite{schwinger1965} There are many other mappings as well,\cite{garbaczewski1978,blaziot1978} including representations based on spin-coherent states.\cite{Radcliffe1971,kuratsuji1980path,Thoss1999mapping,lucke1999semiclassical,song2006coherent}
Each of these mappings is exact on a quantum-mechanical level, but can lead to different levels of approximation when used in a classical trajectory-based simulation.

Among the simplest trajectory-based approaches are quasiclassical methods defined in a mapping basis, which are based on an ensemble of uncoupled trajectories with no phase-dependence on the paths. These can describe electronic coherence effects and have a favourable scaling with system size, at the expense of excluding interference effects in the nuclear dynamics.\cite{miller2012perspective}
One can derive a quasiclassical expression for correlation functions in the MMST basis by linearizing the semiclassical propagator (a method referred to as linearized semiclassical-initial value representation, LSC-IVR).\cite{Miller2001SCIVR,sun1998semiclassical,*Wang1999mapping}
The same dynamics can also be derived by neglecting a cubic term in the quantum-classical Liouville equation (QCLE)\cite{Kapral2006QCL} when written in the MMST basis, an approach called the Poisson-bracket mapping equation (PBME).\cite{Kim2008Liouville,Kelly2012mapping} 
If one goes beyond quasiclassical methods by using the full QCLE or various semiclassical methods, which weight trajectories by phases, the dynamics can also include nuclear interference effects,
which is necessary to predict the correct final momentum distributions in scattering theory.
\cite{Herman+Kluk1984,Stock1997mapping,Sun1997mapping,Bonella2001mapping1}
This is however only feasible for small systems because these approaches get exponentially more difficult as
the system size or simulation time increases.\cite{Stock2005nonadiabatic,Kapral2006QCL}
A possible compromise is provided by partially linearized methods that are somewhere between linearized and semiclassical methods in terms of both accuracy and efficiency. \cite{Huo2011densitymatrix,Hsieh2012FBTS,Hsieh2013FBTS}
Luckily, interference effects become less important in large condensed-phase systems due to fast decoherence.
Our objective in this paper is therefore to find a new quasiclassical method with the same computational cost as the linearized methods mentioned above,
but with an improved accuracy.

The main difference between the MMST and Ehrenfest formulations
is that the former adds a term to the Hamiltonian which can be interpreted as a zero-point energy (ZPE) of the mapping degrees of freedom.
Although this term is rigorously derived,
an unfortunate consequence is that populations can become negative in individual trajectories, which means that the system evolves on inverted potentials.\cite{Bonella2001mapping1,coronado2001,bellonzi2016assessment}
A fundamental problem is that in classical mechanics, this ZPE can flow unphysically between different degrees of freedom, referred to as the ZPE-leakage problem.
M\"{u}ller and Stock have reported that this problem can be mitigated by reducing the ZPE, treating it as an adjustable parameter.\cite{Mueller1999pyrazine} Recently, Cotton and Miller have also used a ZPE-reduced MMST Hamiltonian in their symmetrical quasiclassical windowing approach, finding a particular value to be near optimal for most cases, while still treating it as a parameter that can be tuned for each system or window geometry.\cite{Cotton2013b,Cotton2016}

What MMST and Ehrenfest have in common is that they replace the electronic levels by a quadratic Hamiltonian, whose classical equations of motion are equivalent to the time-dependent Schr\"{o}dinger equation of the subsystem. This idea dates back to Dirac and was further investigated by Strocchi.\cite{dirac1927quantum,strocchi1966complex} However, the quantum phase space of an $N$-level system has a fundamental difference from the classical phase space of a set of $N$ harmonic oscillators.
In quantum mechanics the phase space is described by $N-1$ complex variables (with $2N-2$ real components), 
whereas the classical phase space uses $2N$ free real variables.\cite{zhang1995quantum} The quantum system therefore has one degree of freedom fewer than the classical system. One way to eliminate this degree of freedom from the MMST mapping is via the Holstein-Primakoff transformation,\cite{holstein1940} but this approach is unable to describe Rabi oscillations correctly in semiclassical simulations.\cite{Thoss1999mapping}
Based on Moyal's phase-space theory of quantum mechanics,\cite{moyal1949quantum} Stratonovich formulated a set of general properties to be fulfilled by mappings between quantum and classical mechanics.\cite{stratonovich1957distributions} This approach is now known as the Stratonovich-Weyl transform and can be thought of as a discrete version of the Wigner transform. For an overview of its properties and applications in quantum optics, see Refs.~\onlinecite{varilly1989moyal,klimov2009group,brif1999phase,gadella1995}.

In this paper, we propose a mapping of the two-level system based on the Stratonovich-Weyl transform of a spin-\thalf\ system. This leads to a family of new quasiclassical methods for approximating electronic correlation functions. We show that they are connected with previous approaches by formulating them in terms of an MMST Hamiltonian with reduced ZPE. One of the methods yields the same value of the ZPE parameter as was successfully used by Cotton and Miller.\cite{Cotton2013b} 
We finally investigate the accuracy of our method compared to previous approaches for spin-boson models.

\section{Theory}
Before we present our spin-mapping approach, we briefly revisit previous mappings that are commonly used in quasiclassical calculations. We shall limit the discussion to two states only, although both MMST mapping and Ehrenfest are easily extendable to any finite number of electronic states.

The Hamiltonian of a molecular system with two electronic states in the diabatic representation is
\begin{align} \label{eq:hamiltonian}
    \hat{\mathcal{H}} &= \frac{\hat{p}^2}{2m} + U(\hat{x}) + \hat{V}(\hat{x})  \\
   \hat{V}(x)   &=  \begin{pmatrix} V_1(x) & \Delta^*(x) \\ \Delta(x) & V_2(x) \end{pmatrix},
\end{align}
where $\hat{x}$ and $\hat{p}$ are position and momentum operators of the nuclear modes with associated mass $m$. $U$ and $\hat{V}$ are the state-independent and state-dependent potentials. 
All methods in this paper will employ the classical-path approximation, in which one replaces the nuclear operators by classical phase-space variables, $\hat{x},\hat{p}\mapsto x,p$. What distinguishes the methods is their treatment of the electronic operators and in particular the coupling between the electronic states and the nuclei. Throughout this paper we set $\hbar=1$.

\subsection{MMST mapping and Ehrenfest dynamics} \label{sec:oldmappings}
The so-called Meyer-Miller-Stock-Thoss (MMST) mapping\cite{Meyer1979nonadiabatic,Stock1997mapping} maps the electronic basis states to singly-excited multidimensional harmonic oscillator states
and hence electronic operators to ladder operators
\begin{equation}
    |n\rangle\langle m| \mapsto \hat{a}_n^\dagger \hat{a}_m.
\end{equation}
A transition from state $|m\rangle$ to $|n\rangle$ thus corresponds to moving the single excitation from oscillator $m$ to oscillator $n$. The ladder operators can be written in terms of electronic position and momentum variables as $\hat{a}_n=\frac{1}{\sqrt{2}}(\hat{X}_n+\ii \hat{P}_n)$, so that
\begin{subequations}
\begin{align}
    |n\rangle\langle n| &\mapsto \tfrac{1}{2}(\hat{X}_n^2 + \hat{P}_n^2 - 1) \\
    |n\rangle\langle m|+|m\rangle\langle n| &\mapsto \hat{X}_n \hat{X}_m + \hat{P}_n \hat{P}_m \quad (n\neq m), \\
    \ii(|n\rangle\langle m|-|m\rangle\langle n|) &\mapsto \hat{P}_n \hat{X}_m - \hat{X}_n \hat{P}_m \quad (n\neq m),
\end{align}
\end{subequations}
where we used the commutation relation $[\hat{X}_n,\hat{P}_m]=\ii\delta_{nm}$. Up until this point, the mapping is exact. In the MMST method one now defines a classical limit by removing the hats from the $\hat{X}_n,\hat{P}_n$ operators and treating them as classical canonical variables, which is analogous to the classical-path approximation of the nuclear variables. In other words, electronic and nuclear degrees of freedom are treated on the same footing. The mapped Hamiltonian,
\begin{equation} \label{eq:Ham_MMST}
    H =\frac{p^2}{2m} + U + \sum_{nm} V_{nm}\tfrac{1}{2}(X_n X_m + P_n P_m - \delta_{nm}),
\end{equation}
in which $V_{nm}=\langle n|\hat{V}(x)|m\rangle$, contains a term $\sum_{nm} V_{nm}\frac{1}{2}\delta_{nm}=\frac{1}{2}\mathrm{tr} \,\hat{V}(x)$ that can be interpreted as an electronic zero-point energy (ZPE).\cite{Mueller1999pyrazine}
In this expression we have assumed that $\Delta$ is chosen to be real.

A problem with the MMST method is that in a classical simulation, zero-point energy can flow unrestrictedly between different states, which should not be allowed in a quantum-mechanical simulation.\cite{Stock2005nonadiabatic}
This is referred to as the \emph{ZPE-leakage} problem. 
Stock and M\"{u}ller have found that the ZPE-leakage problem can be mitigated by reducing the overall ZPE.\cite{Mueller1999pyrazine} They introduced a parameter $\gamma$ into the mapped Hamiltonian
\begin{equation} \label{eq:gamma}
    H =\frac{p^2}{2m} + U + \sum_{nm} V_{nm}\tfrac{1}{2}(X_n X_m + P_n P_m - \gamma\delta_{nm})
\end{equation}
which creates a family of methods where $\gamma=1$ corresponds to standard MMST mapping and $\gamma=0$ to Ehrenfest dynamics, as we will see below. Sometimes $\gamma$ is treated as a parameter that is tuned according to some criterion, for example so that average adiabatic populations stay positive.\cite{Mueller1999pyrazine}
In this paper we shall derive a new type of mapping and find that it leads to a Hamiltonian of the form in Eq.~\eqref{eq:gamma}, but a different phase space.

The equations of motion generated by this Hamiltonian are
\begin{subequations}
\begin{equation} \label{eq:eom_electronic}
    \dot{X}_n = \sum_{m}V_{nm} P_m, \qquad \dot{P}_n = -\sum_m V_{nm}X_m,
\end{equation}
\begin{align} 
    \dot{x} &= p/m \\
    \dot{p} &= - \pd{U}{x} - \sum_{nm}  \pd{V_{nm}}{x}\tfrac{1}{2}(X_n X_m + P_n P_m-\gamma\delta_{nm}). \label{eq:eom_nuclear}
\end{align}
\end{subequations}
 From Eq.~\eqref{eq:eom_nuclear}, we see that each potential energy surface, $V_{n}$, contributes to the nuclear force proportionally to
\begin{equation}
    J_{n,\text{cl}} = \tfrac{1}{2}(X_n^2+P_n^2-\gamma)
\end{equation}
which we call the classical population of state $n$. The initial $\{X_n,P_n \}$ variables are typically sampled from either a Wigner distribution or a Dirac delta distribution, depending on the model one wants to simulate. In the first case individual trajectories can have a total population different from one, $J_{1,\text{cl}}+J_{2,\text{cl}}\neq 1$. In both cases populations can even be negative. Negative populations means that the system evolves on inverted potentials, which can lead to problems in particular for systems with steep potentials.\cite{coronado2001,bellonzi2016assessment} It is possible to eliminate $\gamma$ from the equations of motion by choosing a traceless form of $\hat{V}$, so that $\gamma$ only affects the initial distribution and the observables, but does not appear directly in the equations of motion. In other words the MMST mapping is not unique, but depends on a particular choice of splitting between $U$ and $\hat{V}$ (unless the initial conditions are chosen to fix the total population to be one).

A widely used method that gives similar dynamics is the \emph{Ehrenfest} or \emph{mean field trajectory} method,\cite{ehrenfest1927,mclachlan1964} in which one replaces the electronic operators by their expectation values:
\begin{equation}\label{eq:ehrenfest}
    \hat{A} \mapsto \langle \psi| \hat{A}|\psi \rangle.
\end{equation}
Let us write an arbitrary electronic state $|\psi\rangle$ in terms of diabatic basis states as
\begin{equation}
    |\psi\rangle = c_1 |1\rangle + c_2 |2\rangle,
\end{equation}
where $c_n=\frac{1}{\sqrt{2}}(X_n+\ii P_n)$ are the complex coefficients. Note that if $|\psi\rangle$ is normalized and has a fixed global phase, only two of the four real variables $X_1,X_2,P_1,P_2$ are independent (this fact will be used in the next section). Using these coefficients, it is clear that the Ehrenfest Hamiltonian
\begin{equation} \label{eq:Hehrenfest}
\hat{H} \mapsto \frac{p^2}{2m} + U(x) + \sum_{nm} V_{nm}c_n^* c_m 
\end{equation}
is equivalent to the MMST Hamiltonian in Eq.~\eqref{eq:gamma} with $\gamma=0$.\cite{Stock2005nonadiabatic}
The electronic equations of motion~\eqref{eq:eom_electronic} are equivalent to the solution of the time-dependent Schr\"{o}dinger equation for the subsystem
\begin{equation} \label{eq:eom_c}
\dot{c}_n=-\ii \sum_m V_{nm}c_m,
\end{equation}
which is also true for any value of $\gamma$.
Because $\gamma=0$, Ehrenfest dynamics does not suffer from the inverted potential problem. Instead, one finds the \emph{one-trajectory problem}: if the system is initially in state $|1\rangle$, that is $|c_1|^2=\frac{1}{2}(X_1^2+P_1^2)=1$ and $|c_2|^2=\frac{1}{2}(X_2^2+P_2^2)=0$, then for each initial nuclear position and momentum, the dynamics will consist of a single unique trajectory, which on its own cannot capture the correct quantum dynamics. Historically Meyer and Miller solved this problem by introducing a Langer correction,\cite{Meyer1979nonadiabatic} which leads to the MMST Hamiltonian with $\gamma=1$ in Eq.~\eqref{eq:Ham_MMST}. This is not the only possible solution though; any $\gamma\neq 0$ solves the one-trajectory problem.

Throughout this paper, we will repeatedly refer back to the MMST mapping and Ehrenfest dynamics. Before we present our new mapping, let us look at how the Ehrenfest method can be rephrased in terms of a spin-\thalf\ system.

\subsection{Equivalence of a two-level system and a spin-1/2 particle in a magnetic field}\label{sec:spin1/2}
It is well known that the Hamiltonian (and in general any Hermitian operator of the electronic states) can be decomposed into a basis of spin operators and the identity:
\begin{align} \label{eq:Hspinrepr}
    \hat{\mathcal{H}} &= H_0 \hat{\mathcal{I}} + H_x \hat{S}_x + H_y \hat{S}_y + H_z\hat{S}_z \\
    &= H_0 \hat{\mathcal{I}} + \bm{H}\cdot \hat{\bm{S}} \nonumber
\end{align}
where $\hat{S}_i=\frac{1}{2}\hat{\sigma}_i$ ($i\in\{x,y,z\}$) and the Pauli spin matrices are
\begin{equation} \label{eq:spinops}
    \hat{\sigma}_x = \begin{pmatrix} 0 & 1 \\ 1 & 0 \end{pmatrix},~ \hat{\sigma}_y = \begin{pmatrix} 0 & -\ii \\ \ii & 0 \end{pmatrix},~ \hat{\sigma}_z = \begin{pmatrix} 1 & 0 \\ 0 & -1 \end{pmatrix}.
\end{equation}
Such a decomposition was used already in the derivation of the spin-matrix method by Meyer and Miller.\cite{meyer1979spin} The explicit relations between the quantities in Eqs.~\eqref{eq:hamiltonian} and \eqref{eq:Hspinrepr} are
\begin{subequations}
\begin{align}
    H_0 &= \frac{p^2}{2m} + U(x) + \tfrac{1}{2}(V_1(x)+V_2(x)) \\
    H_x &= 2 \,\mathrm{Re} \,\Delta(x) \\
    H_y &= 2 \,\mathrm{Im} \,\Delta(x) \\
    H_z &= V_1(x)-V_2(x).
\end{align}
\end{subequations}
Note that in this way, it does not matter how the initial split between $U$ and $\hat{V}$ is chosen. For simplicity we again assume that $\Delta$ is real, so that $H_y=0$.

The Hamiltonian in equation~\eqref{eq:Hspinrepr} is identical to that of a spin-\thalf\ particle in a magnetic field $-\bm{H}$ (if we choose its gyromagnetic ratio to be one). We emphasize that this spin is only a theoretical tool and has no relation to any real spins of the physical system. 
Since the Hamiltonians in Eqs.~\eqref{eq:hamiltonian} and \eqref{eq:Hspinrepr} are in one-to-one correspondence, any classical phase space of spin-\thalf\ particles will also be a phase space of the general two-level system.

\begin{figure}
\centering
\resizebox{0.6\columnwidth}{!}{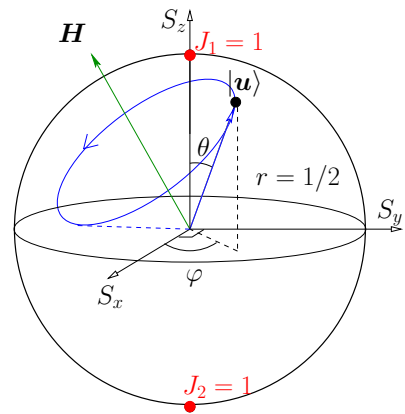}~
\caption{The expectation values of the spin operators form the Cartesian coordinates of a sphere with radius 1/2. Their equations of motion describe precession of the spin vector around the external magnetic field, which is in one-to-one correspondence with the two-level potential-energy matrix. The ``north'' and ``south'' poles indicate points where all population is concentrated in one of the basis states of the subsystem.}\label{fig:Bsphere}
\end{figure}

In order to construct a classical phase space, we use spin coherent states defined as\cite{Radcliffe1971}
\begin{equation}
    |\bm{u}\rangle = \cos\tfrac{\theta}{2}\eu{-\ii\varphi/2}|1\rangle + \sin\tfrac{\theta}{2}\eu{\ii\varphi/2}|2\rangle,
\end{equation}
where $\bm{u}$ denotes a unit vector defined by the spherical coordinates $\theta,\varphi$ and the states are normalized such that $\braket{\bm{u}|\bm{u}}=1$.
Note that this defines an arbitrary state using two real variables, in contrast with the MMST mapping that is defined with four real variables. The expectation values of the spin operators $\hat{S}_i$ in this state are 
\begin{equation} 
S_i(\bm{u}) \equiv \langle \bm{u}|\hat{S}_i|\bm{u}\rangle = \tfrac{1}{2}u_i ,
\end{equation}
where
\begin{subequations}\label{eq:Sxyz}
\begin{align}
    u_x &= \sin\theta \cos\varphi  \\
    u_y &= \sin\theta \sin\varphi  \\
    u_z &= \cos\theta.
\end{align}
\end{subequations}
Thus $S_i(\bm{u})$ are the Cartesian coordinates of a sphere with radius $r=1/2$, see Figure~\ref{fig:Bsphere}. One can think of them as orthogonal functions analogous to p-orbitals. Likewise the populations are
\begin{subequations}
\begin{align} 
J_1(\bm{u})&\equiv \langle \bm{u}|1\rangle\langle 1|\bm{u}\rangle = \cos^2\tfrac{\theta}{2} \\ 
J_2(\bm{u})&\equiv \langle \bm{u}|2\rangle\langle 2|\bm{u}\rangle =\sin^2\tfrac{\theta}{2}.
\end{align}
\end{subequations}
and it is clear that $J_1(\bm{u})+J_2(\bm{u})=1$ is always fulfilled.
Note that on the ``poles'' of the sphere along the $S_z$-axis ($\theta=0,\pi$), all population is concentrated in one of the electronic basis states, as shown in Figure~\ref{fig:Bsphere}.

Let us now briefly discuss the dynamics of the spin. It is known from quantum mechanics that the vector of spin operators follows the (Heisenberg) equation of motion
\begin{equation}
    \frac{\rd}{\rd t}\hat{\bm{S}} = \bm{H} \times \hat{\bm{S}}.
\end{equation}
Replacing all operators with their expectation values $\bm{S}=(S_x(\bm{u}),S_y(\bm{u}),S_z(\bm{u}))$ gives
\begin{equation} \label{eq:eom_spin}
    \frac{\rd}{\rd t}\bm{S}= \bm{H} \times \bm{S},
\end{equation}
which describes precession of the spin vector around the magnetic field, see Figure~\ref{fig:Bsphere}. 
These equations of motion are also equivalent to the time-dependent Schr\"{o}dinger equation of the subsystem in Eq.~\eqref{eq:eom_c}.\footnote{This can be shown by setting $2S_x=\text{Re}[c_1^*c_2]$, $2S_y=\text{Im}[c_1^*c_2]$, $2S_z=\thalf (|c_1|^2-|c_2|^2)$.} The dynamics preserves the magnitude $|\bm{S}|$, i.e. the radius of the sphere. It is the $x$-component of the field, $H_x=2\Delta$, that drives population transfer between the two states, whereas the $z$-component, $H_z=V_1-V_2$, preserves all populations and only contributes with a relative phase.

What we have presented in this section is an alternative formulation of the subsystem dynamics in the Ehrenfest method, which also appears in Ref.~\onlinecite{meyer1979spin}. In the next section, we will generalize the above procedure to allow for a phase-space representation of correlation functions,
leading to an accurate quasiclassical methodology.

\subsection{Stratonovich-Weyl transform of the spin-1/2 system} 
In the previous section we replaced spin operators by their expectation values. Let us for a general operator $\hat{A}$ associate this procedure with the mapping
\begin{equation} \label{eq:Q}
\hat{A}\mapsto A_\text{Q}(\bm{u}) \equiv \langle \bm{u} |\hat{A}|\bm{u} \rangle, 
\end{equation}
which in literature is known as the Q-function.\cite{klimov2009group} The Q-function allows us evaluate a quantum-mechanical trace over the subsystem, $\tr[\hat{A}]$, as an integral over a classical phase space:
\begin{equation}\label{eq:trA}
    \tr[\hat{A}] = \int \rd\bm{u} \, A_\text{Q}(\bm{u}) = \frac{1}{2\pi}\int_0^{2\pi} \rd\varphi \int_0^\pi \rd \theta \sin\theta A_\text{Q}(\bm{u}),
\end{equation}
where the integration measure is $\rd\bm{u} = \tfrac{1}{2\pi}\rd\varphi\rd\theta\sin\theta $ (which is normalized to be consistent with $\tr[\hat{\mathcal{I}}]=2$).
However, in order to calculate correlation functions, we need a phase-space representation for \emph{products} of operators of the form $\tr[\hat{A}\hat{B}]$. The Q-function is not enough for this purpose since
\begin{equation} 
    \tr[\hat{A}\hat{B}] \neq \int \rd\bm{u}\, A_\text{Q}(\bm{u})B_\text{Q}(\bm{u}),
\end{equation}
because of the uncertainty property $\langle \bm{u}| \hat{A}\hat{B}|\bm{u}\rangle \neq \langle\bm{u}| \hat{A}|\bm{u}\rangle \langle\bm{u}| \hat{B}|\bm{u}\rangle$. The only case when equality holds is when $\hat{A}$ or $\hat{B}$ is (proportional to) the identity operator $\hat{\mathcal{I}}$.
This indicates that we need to treat the spin operators differently
from the identity. Note that the idea of such a separation was recently investigated in the MMST representation.\cite{saller2019} Now we will show how the uncertainty property of the spin operators can be included directly in the construction of the phase space. First, note that Eq.~\eqref{eq:Q} can written equivalently as the result of the quantum-mechanical trace:
\begin{equation}\label{eq:Q2}
    A_\text{Q}(\bm{u}) \equiv \tr[\hat{A}\hat{w}_\text{Q}(\bm{u})], \quad  \hat{w}_\text{Q}(\bm{u}) = \frac{1}{2}\hat{\mathcal{I}} + \frac{1}{2}\bm{u}\cdot \hat{\bm{\sigma}},
\end{equation}
Next, introduce the P-function\footnote{The Q- and P-functions are sometimes called Berezin's covariant and contravariant symbols \cite{berezin1975general}}
\nocite{berezin1975general}
\begin{equation} \label{eq:P}
    A_\text{P}(\bm{u}) \equiv \tr[\hat{A}\hat{w}_\text{P}(\bm{u})], \quad \hat{w}_\text{P}(\bm{u}) = \frac{1}{2}\hat{\mathcal{I}} + \frac{3}{2}\bm{u}\cdot \hat{\bm{\sigma}},
\end{equation}
which together with the Q-function allows us to represent $\tr[\hat{A}\hat{B}]$ as
\begin{equation} \label{eq:trAB}
    \tr[\hat{A}\hat{B}] = \int \rd\bm{u} \, A_\text{Q}(\bm{u})B_\text{P}(\bm{u}) = \int \rd\bm{u} \, A_\text{P}(\bm{u})B_\text{Q}(\bm{u}).
\end{equation}
Interestingly, one can also define a self-dual W-function
\begin{equation} \label{eq:W}
     A_\text{W}(\bm{u}) \equiv \tr[\hat{A}\hat{w}_\text{W}(\bm{u})], \quad \hat{w}_\text{W}(\bm{u}) = \frac{1}{2}\hat{\mathcal{I}} + \frac{\sqrt{3}}{2}\bm{u}\cdot \hat{\bm{\sigma}},
\end{equation}
which fulfils
\begin{equation} \label{eq:trAB_W}
    \tr[\hat{A}\hat{B}] = \int \rd\bm{u} \, A_\text{W}(\bm{u})B_\text{W}(\bm{u}).
\end{equation}
We will refer to the formulas~\eqref{eq:trAB} and \eqref{eq:trAB_W} as the \emph{traciality} property of the mapping.
It is easy to show (using the special case $\hat{B}=\hat{\mathcal{I}}$) that the P- and W-functions also fulfil equations of the form of Eq.~\eqref{eq:trA}.

It is natural to think of the mapping to the phase-space representation in Eq.~\eqref{eq:W}, and the pair of Eqs.~\eqref{eq:Q2} and~\eqref{eq:P}, as a discrete version of the Wigner transform. In literature this is known as the Stratonovich-Weyl transform.\cite{stratonovich1957distributions,brif1999phase} It has been applied to various problems in quantum optics\cite{klimov2009group} but to our knowledge not yet in the context of nonadiabatic molecular dynamics.
To summarize, let us write the \emph{Stratonovich-Weyl kernels} $\hat{w}_s(\bm{u})$ ($s\in\{\text{Q},\text{P},\text{W}\}$) collectively as
\begin{equation}
    \hat{w}_s(\bm{u}) = \tfrac{1}{2}\hat{\mathcal{I}} + r_s\bm{u}\cdot\hat{\bm{\sigma}}
\end{equation}
where we introduced the \emph{spin radius} $r_s$, defined in Table~\ref{tab:symbols} for each $s$. As simple examples, we have
\begin{equation}
    [\hat{\mathcal{I}}]_s(\bm{u}) = 1 \quad \text{and} \quad [\hat{S}_i]_s(\bm{u}) = r_s u_i.
\end{equation}
and all other operators can be built as linear combinations of these.
In other words, the Q-function of the quantum spin vector $\hat{\bm{S}}$ defines the same sphere with radius $r_\text{Q}=1/2$ that we saw in Figure~\ref{fig:Bsphere}, whereas the corresponding W- and P-functions define larger spheres of radii $r_\text{W}=\sqrt{3}/2$ and $r_\text{P}=3/2$, respectively. 

\begin{table}
    \centering
    \caption{Spin radii, $r_s$, and dual symbols, ${\bar{s}}$, of the three phase-space functions.}
    \label{tab:symbols}
    \begin{ruledtabular}
    \begin{tabular}{lccc}
        $s$ & Q & W  & P  \\ \hline
        ${\bar{s}}$ & P & W & Q \\
        $r_s$ & 1/2 & $\sqrt{3}/2$ & 3/2 \\ 
    \end{tabular}
    \end{ruledtabular}
\end{table}

\begin{figure}
    \centering
    \resizebox{0.6\columnwidth}{!}{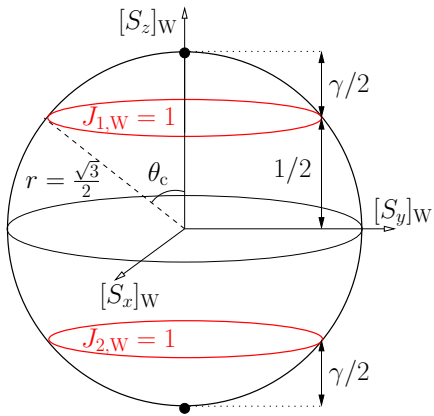}~
    \caption{The spin operators in the W-representation form the Cartesian components of a sphere with radius $\sqrt{3}/2$. The system is entirely in $|1\rangle$ at the northern arctic circle defined by $\cos\theta_\text{c}=1/\sqrt{3}$, instead of at the north pole as in Figure~\ref{fig:Bsphere}. For $0< \theta<\theta_\text{c}$ we have $J_1>1$ and $J_2<0$, and vice versa for $\pi-\theta_\text{c}<\theta<\pi$. The ZPE parameter $\gamma=\sqrt{3}-1$ measures the size of the regions with negative populations. The corresponding P-sphere looks similar but with a larger radius $3/2$ and $\gamma=2$, meaning that the red circles lie closer to the equator ($\cos\theta_\text{c}=1/3$).
    }
    \label{fig:arctic_circles}
\end{figure}

In quantum mechanics, the squared magnitude of the spin is
\begin{equation}
    \hat{\bm{S}}^2 \equiv \hat{S}_x^2+ \hat{S}_y^2 + \hat{S}_z^2 = \frac{3}{4}\hat{\mathcal{I}} = \frac{1}{2}\left(\frac{1}{2}+1\right)\hat{\mathcal{I}}.
\end{equation}
Its eigenvalues are equal to $r_\text{W}^2$, i.e.\ $r_\text{W}$ is the quantum magnitude of the spin vector.
Thus the W-sphere in Figure~\ref{fig:arctic_circles} can be interpreted as a classical phase space of spin. Also the Q- and P-functions give the correct quantum magnitude of spin as long as they are used in pairs ($r_\text{Q}r_\text{P}=3/4$) as in Eq.~\eqref{eq:trAB}.

Let us now take a step back and see what consequences this has for the general two-level system. 
Populations are in the spin-mapping space represented as 
\begin{subequations}
\begin{align}
    |1\rangle\langle 1|&=\tfrac{1}{2}\hat{\mathcal{I}}+\hat{S}_z \mapsto \tfrac{1}{2} + r_s\cos\theta\equiv J_{1,s},\\
    |2\rangle\langle 2|&=\tfrac{1}{2}\hat{\mathcal{I}}-\hat{S}_z \mapsto \tfrac{1}{2} - r_s\cos\theta\equiv J_{2,s},
\end{align}
\end{subequations}
and $J_{1,s}+J_{2,s}=1$ for all $s$.
As a result, the points on the phase-space sphere where all population is concentrated to one eigenstate are no longer on the poles but are identified with ``polar circles'' defined by $\cos\theta_\text{c}=\pm 1/(2r_s)$, as indicated for the W-function in Figure~\ref{fig:arctic_circles}. In the region ``north'' of the ``arctic'' circle one finds $J_{1,s}>1$ and $J_{2,s}<0$, and vice versa to the ``south'' of the ``antarctic'' circle. In the Q-case, $J_{n,\mathrm{Q}}=1$ only on the poles as in Figure~\ref{fig:Bsphere}, so that the are no points with negative populations, and in the P-case the situation is similar to in Figure~\ref{fig:arctic_circles}, but the circles with $J_{n,\text{P}}=1$ are closer to the equator, like tropical circles.

This has consequences for the dynamics in the three spin-mapping representations. The Hamiltonian in equation~\eqref{eq:Hspinrepr} maps to
\begin{align}\label{eq:Hs}
    H_s(\bm{u}) &= H_0 + r_s \bm{H}\cdot \bm{u} \\
    &=\frac{p^2}{2m} + U + V_1 J_{1,s} + V_2 J_{2,s} + 2r_s \Delta \sin\theta\cos\varphi. \nonumber
\end{align}

\noindent In Appendix~\ref{sec:app_QCLE} we derive the equations of motion using a spin-mapping representation of the QCLE (within an uncoupled-trajectory approximation), which allows us describe the dynamics of both the subsystem and the environment.\cite{Kapral2006QCL} The resulting equations of motion are
\begin{subequations}\label{eq:eom_all}
\begin{align}
    \dot{\bm{u}} &= \bm{H} \times \bm{u} \label{eq:eom_n} \\
    \dot{x} &= p/m \\
    \dot{p} &= -\pd{H_0}{x}-r_s \pd{\bm{H}}{x}\cdot\bm{u},
\end{align}
\end{subequations}
which can be shown to conserve $H_s(\bm{u})$. 
Importantly the subsystem equations are exact and equal for all $s$, and preserve the magnitude $|\bm{u}|$.
Thus the only dependence on $s$ is in the nuclear equations of motion. Firstly, the contribution from the electronic coupling, $\Delta$, scales by $r_s$. Secondly, the region of phase space corresponding to inverted potential have different size. In this aspect, $H_\text{Q}$ is interesting because it never gives inverted potentials (it is the same Hamiltonian as is used in Ehrenfest dynamics). The W-representation on the other hand has the attractive property of being self-dual.

\subsection{Comparison with previous methods}
Let us now connect spin mapping to the previous methods introduced in section~\ref{sec:oldmappings}.
We introduce the coordinate transformation
\begin{subequations} \label{eq:ntoXP}
\begin{align}
    2r_s u_x &= X_1 X_2 + P_1 P_2 \equiv \tilde{\sigma}_x(\mathsf{X},\mathsf{P}) \\
    2r_s u_y &= X_1 P_2 - X_2 P_1  \equiv \tilde{\sigma}_y(\mathsf{X},\mathsf{P}) \\
    2r_s u_z &= \tfrac{1}{2}(X_1^2 + P_1^2 - X_2^2 -P_2^2)  \equiv \tilde{\sigma}_z(\mathsf{X},\mathsf{P}),
\end{align}
\end{subequations}
where $\bm{\tilde{\sigma}}=(\tilde{\sigma}_x,\tilde{\sigma}_y,\tilde{\sigma}_z)$
is the MMST representation of the Pauli matrices.
Using $|\bm{u}|^2=1$ one can show that 
\begin{equation}\label{eq:XPradius}
    4r_s= X_1^2+P_1^2+X_2^2+P_2^2 \equiv R^2,
\end{equation}
i.e.\ the new variables are also constrained to a sphere (a 3-dimensional hypersphere embedded in $\mathbb{R}^4$) but with a different radius $R=\sqrt{4r_s}$. The transformation used together with Eq.~\eqref{eq:XPradius} brings the Hamiltonian in Eqs.~\eqref{eq:Hs} to the form
\begin{multline}
    H_s = \frac{p^2}{2m} + U + \sum_{n=1}^2 V_n\tfrac{1}{2}(X_n^2+P_n^2-2r_s+1) \\ 
    + \Delta (X_1X_2+P_1P_2).
\end{multline}
If $\{X_n,P_n\}$ are identified as canonical variables of the subsystem, we get equations of motion equivalent to Eq~\eqref{eq:eom_all}, and $R$ is constant.
Therefore this Hamiltonian is equivalent to the ZPE-reduced MMST Hamiltonian in equation~\eqref{eq:gamma} with $\gamma=2r_s-1$. In particular $H_\text{W}$ has $\gamma=\sqrt{3}-1\approx 0.732$, which is the value Cotton and Miller gave a heuristic argument for in Ref.~\onlinecite{Cotton2013b}, also based on an analogy with spin (note that they define $\gamma$ as one half times our definition). It is also close to values that M\"{u}ller and Stock found optimal to reproduce the correct level density of various models,\cite{Mueller1999pyrazine} and Golosov and Reichman have showed that it gives correct short-time dynamics up to seventh order in time for some observables, compared to fifth order with standard MMST mapping.\cite{golosov2001mapping}
Note however that in both the last two references, the mapping Hamiltonian is defined as $\sum_{nm} H_{nm}\tfrac{1}{2}(X_nX_m+P_nP_m-\gamma\delta_{nm})$ where $H_{nm}=(\tfrac{p^2}{2m}+U)\delta_{nm}+V_{nm}$. This is different from our approach, which treats terms proportional to the identity differently than other operators.



The split between $U$ and $\hat{V}$ is therefore unambiguous in our approach. This is clear from Eq.~\eqref{eq:Hs} that only multiplies the traceless part of $\hat{V}$ by $r_s$. In this way, when $\hat{V}$ goes to $\hat{\mathcal{I}}$, the mapped Hamiltonian smoothly reduces to the single-state Hamiltonian.
In LSC-IVR however, different choices of the split between $U$ and $\hat{V}$ generally lead to different results since the total population $J_{1,\text{cl}}+J_{2,\text{cl}}$ is allowed to deviate from one. A further difference is that Wigner initial conditions in MMST mapping corresponds to a Gaussian distribution in four dimensions with unbounded variables.
In our case, the phase-space variables are bounded by the constraint to a sphere.
This leads us to identify the most important difference: since the entire spin-mapping space is isomorphic to the physical space, there is no need to introduce projections onto a physical subspace as in the MMST mapping.

Compared to another recent spin mapping by Cotton and Miller,\cite{Cotton2015} our method maps two levels to \emph{one} spin-\thalf\ particle, whereas Cotton and Miller map to \emph{two} spin-\thalf\ particles. Thus their mapping space has, just like in standard MMST mapping, more degrees of freedom than the underlying quantum system.

A link between the MMST mapping and spin coherent states was found already by Stock and Thoss,\cite{Thoss1999mapping}
who related their work to the spin-coherent state propagator by Suzuki.\cite{suzuki1983classical}
They expressed a semiclassical propagator in the MMST representation as an integral over the coordinates of the spin coherent state\footnote{Stock and Thoss used a complex coherent state variable $\mu$ which is effectively a stereographic projection of our $\bm{u}$ vector onto a complex plane.
} and what is in our notation the radius $R$.
They then performed the approximation of fixing $R$ in order to relate it to Suzuki's propagator. In our work we directly derive the spin coherent state representation, which naturally leads to a fixed $R$, and additionally opens up the path towards defining correlation functions via the Q-, P- and W-functions, as we describe in the next section. 

\subsection{Correlation functions from the various methods}
We will now use our phase-space representation from the previous sections to approximate correlation functions of the type
\begin{equation}
    C_{AB}(t) = \mathrm{Tr}[\hat{\rho}\hat{A}(0)\hat{B}(t)],
\end{equation}
where $\Tr$ denotes a full quantum-mechanical trace, $\hat{\rho}$ is a density matrix normalized to give $\Tr[\hat{\rho}]=1$, and $\hat{B}(t)=\eu{\ii\hat{H}t}\hat{B}\eu{-\ii\hat{H}t}$. In this paper we will only consider the case when $\hat{A},\hat{B}$ are electronic operators and the density matrix is state-independent,
$\hat{\rho}=\hat{\rho}_\text{nuc}\otimes\hat{\rho}_\text{el}$, where
$\hat{\rho}_\text{el}=\frac{1}{2}\hat{\mathcal{I}}$ (the normalization ensures that $\tr[\hat{\rho}_\text{el}]=1$). 
The initial projection onto one of the electronic states is defined by $\hat{A}$.

We will use a Wigner distribution of the $F$ nuclear degrees of freedom,
\begin{equation}
    \rho_\text{nuc}(x,p) = \frac{1}{(2\pi)^F}\int \eu{\ii py}\left\langle x-\frac{y}{2}\right|\hat{\rho}_\text{nuc} \left|x+\frac{y}{2}\right\rangle \rd y
\end{equation}
and a quasiclassical propagator analogous to classical Wigner dynamics:
\begin{equation} \label{eq:CAB}
C_{AB}(t) \approx  \int \rd x \, \rd p \frac{\rd\bm{u}}{2} \, \rho_\text{nuc}(x,p) A_{s}(\bm{u}) B_{\bar{s}}(\bm{u}(t)),
\end{equation}
where the dual indices ${\bar{s}}$ are specified in Table~\ref{tab:symbols}. The trajectory $\bm{u}(t)$ is generated by $H_s$, with the same index as the operator at time zero.
(We also considered the opposite index choice $A_{\bar{s}}B_s$ but found that it gives less accurate results at $t>0$.)

We integrate over $\frac{\rd\bm{u}}{2}=\frac{\rd\varphi}{4\pi}\rd\theta\sin\theta$ using a Monte Carlo scheme, drawing samples uniformly from the unit sphere $|\bm{u}|^2=1$. Using Eq.~\eqref{eq:XPradius}, this can be done directly in the $\{X_n,P_n\}$ variables by sampling from a 3-sphere with radius $R=\sqrt{4r_s}=\sqrt{2(\gamma+1)}$. An easy way to do this in practice is to sample $\{X_n,P_n\}$ independently from a standard normal distribution and multiply them with a common factor to fulfil Eq.~\eqref{eq:XPradius}.

Eq.~\eqref{eq:CAB} gives us three new methods classified by $s$: we will call them the Q-, P- and W-methods. To evaluate their performance, we will compare them to two standard quasiclassical methods in the MMST-basis. These differ in how the system is projected onto the physical subspace:\cite{Kelly2012mapping}
the first, LSC-IVR, has projection operators both at initial and final times,\cite{Miller2001SCIVR,sun1998semiclassical}
and the second, PBME, has a projection only at initial times.\cite{Kim2008Liouville}
As a general formula we write 
\begin{multline} \label{eq:CABgeneral}
    C_{AB}(t) \approx \int \rd x\,\rd p\,\rd \mathsf{X}\,\rd\mathsf{P}\, \rho_\text{nuc}(x,p) \rho_\text{el}(\mathsf{X},\mathsf{P}) \\ \times A(\mathsf{X},\mathsf{P})B(\mathsf{X}(t),\mathsf{P}(t)),
\end{multline}
where $\mathsf{X}=(X_1,X_2)$, $\mathsf{P}=(P_1,P_2)$. The explicit expressions for $\rho_\text{el}(\mathsf{X},\mathsf{P})$ are given in Table~\ref{tab:methods} together with the appropriate observables in the various methods.
For a general operator $\hat{A} = A_0\hat{\mathcal{I}} + \thalf\bm{A}\cdot\hat{\bm{\sigma}}$, the MMST mapping representations 
used in the standard formulations of LSC-IVR and PBME are defined as
\begin{align}
    A_\text{wig} &= \thalf (R^2-2) A_0 + \thalf \bm{A}\cdot\tilde{\bm{\sigma}}
    \\
    A_\text{SEO} &= \tilde{A}_\text{SEO} \phi
    \\
    \tilde{A}_\text{SEO} &= \thalf (R^2-1) A_0 + \thalf \bm{A}\cdot\tilde{\bm{\sigma}},
\end{align}
where $A_\text{SEO}$ is used whenever one projects the harmonic oscillators to the physical subspace, and $\phi=16\exp(-R^2)$.
The special case of a population operator, $\hat{A}=\ketbra{n}{n}=(\hat{\mathcal{I}}+\hat{\sigma}_z)/2$ gives
\begin{subequations}\label{eq:mmstops}
\begin{align}
    A_\text{wig}  &= \tfrac{1}{2}(X_n^2 + P_n^2-1)  \\
    A_\text{SEO}  &= \tfrac{1}{2}\left(X_n^2+P_n^2 - \tfrac{1}{2}\right)\phi \equiv \tilde{A}_\text{SEO}\phi.
\end{align}
\end{subequations}

\noindent Similarly we can also write the Stratonovich-Weyl observables in terms of $\{\mathsf{X},\mathsf{P}\}$ using Eq.~\eqref{eq:ntoXP}, which gives
\begin{align*}
    A_{\text{SW}} &= A_0 + r_{s} \bm{A} \cdot \bm{u}
    = A_0 + \thalf \bm{A} \cdot \bm{\tilde{\sigma}}(\mathsf{X},\mathsf{P})
    \\
    B_{\overline{\text{SW}}} &= B_0 + r_{\bar{s}} \bm{B} \cdot \bm{u} 
    = B_0 + \frac{3}{2(\gamma+1)^2} \bm{B} \cdot \bm{\tilde{\sigma}}(\mathsf{X},\mathsf{P}).
\end{align*}
Note that $A_\text{SW}$ gives a standard expression for a population operator, $A_\text{SW} = \thalf (X_n^2+P_n^2-\gamma)=J_{n,\text{cl}}$, whereas $B_{\overline{SW}}$ has not previously been used in quasiclassical mapping dynamics.



All methods presented so far treat initial conditions via a weighting procedure. We will contrast these to methods that use focused initial conditions, which are used in Ehrenfest dynamics and sometimes also with the MMST mapping. By focused we mean that to start in for example $\hat{A} = |1\rangle\langle 1|$, an Ehrenfest calculation would use the initial conditions $|c_1|^2=1$ and $|c_2|^2=0$, which in the MMST representation is
\begin{equation}\label{eq:focused}
    \rho_\text{el} A = \mathcal{N}\delta(R^2-2(\gamma+1)) \, \delta(X_1^2+P_1^2-(\gamma+2)),
\end{equation}
where $\mathcal{N}=\pi^{-2}$ (although the normalization is not explicitly needed when evaluating with a simple Monte Carlo procedure).
In a similar way,
we define a focused initial condition for the W-representation by sampling $\bm{u}$ from the northern polar circle in Figure~\ref{fig:arctic_circles}, which corresponds to Eq.~\eqref{eq:focused} with $\gamma=\sqrt{3}-1\approx 0.732$.
A focused Q-distribution is identical to Ehrenfest dynamics. Unlike the approaches in Table~\ref{tab:methods}, we cannot derive the focused methods in a rigorous way from the Stratonovich-Weyl formalism
but instead follow the same line of reasoning as Refs.~\onlinecite{Bonella2003mapping,Mueller1999pyrazine}.

\begin{table}[tb]
    \caption{Definition of the terms used in Eq.~\eqref{eq:CABgeneral} to compute general correlation functions $C_{AB}(t)$ for various methods.
    We use the short-hand notation $R^2=X_1^2+P_1^2+X_2^2+P_2^2$, $\phi=16\exp(-R^2)$ and $\mathcal{N}=[2\pi^2(\gamma+1)]^{-1}$.}
    \label{tab:methods}
    \centering
    \begin{ruledtabular}
    \begin{tabular}{lccccc}
        Method & $\rho_\text{el}(\mathsf{X},\mathsf{P})$ & $\gamma$ & $A$ & $B$ \\ \hline
        Q & $\mathcal{N}\delta(R^2-2(\gamma+1))$ & 0 & $A_\text{SW}$ & $B_{\overline{\text{SW}}}$ \\
        W & $\mathcal{N}\delta(R^2-2(\gamma+1))$ & $\sqrt{3}-1$ & $A_{\text{SW}}$ & $B_{\overline{\text{SW}}}$ \\
        P & $\mathcal{N}\delta(R^2-2(\gamma+1))$ & 2 & $A_\text{SW}$ & $B_{\overline{\text{SW}}}$ \\
        PBME & $\phi/(8\pi^2)$ & 1 & $\tilde{A}_\text{SEO}$ & $B_\text{wig}$ \\ 
        LSC-IVR & $\phi^2/(8\pi^2)$ & 1  & $\tilde{A}_\text{SEO}$ & $\tilde{B}_\text{SEO}$ \\ 
    \end{tabular}
    \end{ruledtabular}
\end{table}

\begin{table}[tb]
    \caption{Definition of the terms used in Eq.~\eqref{eq:CABgeneral} to compute correlation functions $C_{AB}(t)$,
    where the initial state is $\hat{A}= |n\rangle\langle n|$,
    for various methods employing focused initial conditions.
    The electronic distribution in each case is given by
    $\rho_\text{el}(\mathsf{X},\mathsf{P})=\mathcal{N}\delta(R^2-2(\gamma+1))$ and $A=\delta(X_n^2+P_n^2-(\gamma+2))$, with $\mathcal{N}=\pi^{-2}$.
    }
    \label{tab:focused}
    \centering
    \begin{ruledtabular}
    \begin{tabular}{lcc}
        Method &  $\gamma$ & $B$ \\ \hline
        Ehrenfest  & 0 & $B_\text{SW}=B_\text{MFT}$ \\
        MMST-focused & $1$ & $B_\text{SW}=B_\text{wig}$   \\
        W-focused  & $\sqrt{3}-1$ & $B_\text{SW}$ \\        
    \end{tabular}
    \end{ruledtabular}
\end{table}

Note that when using focused initial conditions, the Stratonovich-Weyl operators reduce to the usual Ehrenfest (MFT) and MMST operators. For $\gamma=0$
\begin{align}
    B_\text{SW} 
    &= \tfrac{1}{2} \left[ R^2 B_0 + \bm{B} \cdot \bm{\tilde{\sigma}}(\mathsf{X},\mathsf{P}) \right]
    \\
    &= \sum_{nm} B_{nm} c_n^* c_m \equiv B_\text{MFT},
\end{align}
and similarly for $\gamma=1$ one has $1=(R^2-2)/2$, so that $B_\text{SW}=B_\text{wig}$.

%
%


\section{Results and discussion}
To test the accuracy of our method, we have applied it on a spin-boson model,\cite{Garg1985spinboson} which is a standard benchmark problem consisting of two electronic levels coupled to a harmonic bath of nuclear modes. Note however that all quasiclassical mapping approaches can be applied to more general problems with anharmonic potentials.
The potentials defined in Eq.~\eqref{eq:hamiltonian} are of the form
\begin{align}
    U(x) &= \sum_{j=1}^F \tfrac{1}{2}m_j\omega_j^2 x_j^2, \\
    \hat{V}(x) &= \Delta \hat{\sigma}_x + \left(\varepsilon+\sum_{j=1}^F c_j x_j\right) \hat{\sigma}_z.
\end{align}
Here $\varepsilon$ is the energy bias and $\Delta$ the constant diabatic coupling between the two electronic states. The bath consists of $F$ nuclear modes, each with frequency $\omega_j$ and vibronic coupling coefficient $c_j$. In this form the matrix $\hat{V}$ is already traceless.  We choose units such that $\hbar=1$.

The so-called spectral density of the bath, $J_\text{bath}(\omega)$, determines the distribution of nuclear frequencies. Among the most common forms of this function is the Ohmic bath
\begin{equation}
    J_\text{bath}(\omega) = \frac{\pi\xi}{2} \omega\, \mathrm{e}^{-\omega/\omega_\text{c}},
\end{equation}
where $\omega_\text{c}$ is called the characteristic frequency and $\xi$ the Kondo parameter,
which determines the strength of the friction.
Numerically one uses a discretization
\begin{equation}
    J_\text{bath}(\omega) = \frac{\pi}{2}\sum_{j=1}^F \frac{c_j^2}{m_j\omega_j}\delta(\omega-\omega_j)
\end{equation}
with a finite $F$. In this work we used the discretization scheme employed in Ref.~\onlinecite{craig2005rate}.

To sample the initial nuclear variables $\{x_j,\;p_j\}$, we used the thermal Wigner distribution
\begin{multline}
    \rho_\text{nuc}(x,p) = \prod_{j=1}^F \frac{\alpha_j}{\pi} \exp\left[-\frac{2\alpha_j}{\omega_j}\left(\frac{p_j^2}{2m_j}+\frac{1}{2}m_j\omega_j^2x_j^2\right)\right],
\end{multline}
with $\alpha_j=\tanh\left(\frac{1}{2}\beta\omega_j\right)$ and $\beta=1/k_\text{B}T$.

To facilitate comparison with other recent papers, we have used the same parameter settings as in Ref.~\onlinecite{Cotton2016}, which are given in Table~\ref{tab:parameters}. Note that we define our initial nuclear distribution to be centred around the minimum of $U$, which for these systems gives practically identical results to when starting around $U+V_1$ as in Ref.~\onlinecite{Cotton2016} (except for subtle changes in the strong-bath case). In our calculations we used $F=100$ nuclear modes and a timestep of $0.01\Delta^{-1}$.
To ensure very tight convergence, we used $10^5$ trajectories
for the methods with focused initial conditions and $10^6$ trajectories for the others.
However, like other mapping approaches, $10^3-10^4$ is typically enough to observe the correct physical behaviour with only a little numerical noise.

\begin{table}
    \centering
    \caption{Parameter settings of the spin-boson models, referring to the panels in Figures~\ref{fig:center}--\ref{fig:center_foc}.}
    \label{tab:parameters}
    \begin{ruledtabular}
    \begin{tabular}{lcd{1.2}d{1.2}c}
        Panel&$\varepsilon/\Delta$ &\head{$\xi$}   &\head{$\beta\Delta$}  &$\omega_\text{c}/\Delta$  \\ \hline
        (a)   & 0             & 0.09      & 0.1           & 2.5 \\
        (b)   & 0             & 0.09      & 5             & 2.5 \\
        (c)   & 1             & 0.1       & 5             & 2.5 \\
        (d)   & 1             & 0.1       & 0.25          & 1.0 \\
        (e)   & 0             & 2         & 1             & 1.0 \\
        (f)   & 5             & 4         & 0.1           & 2.0 \\ 
    \end{tabular}
    \end{ruledtabular}
\end{table}

\begin{figure*}
    \centering
    \includegraphics{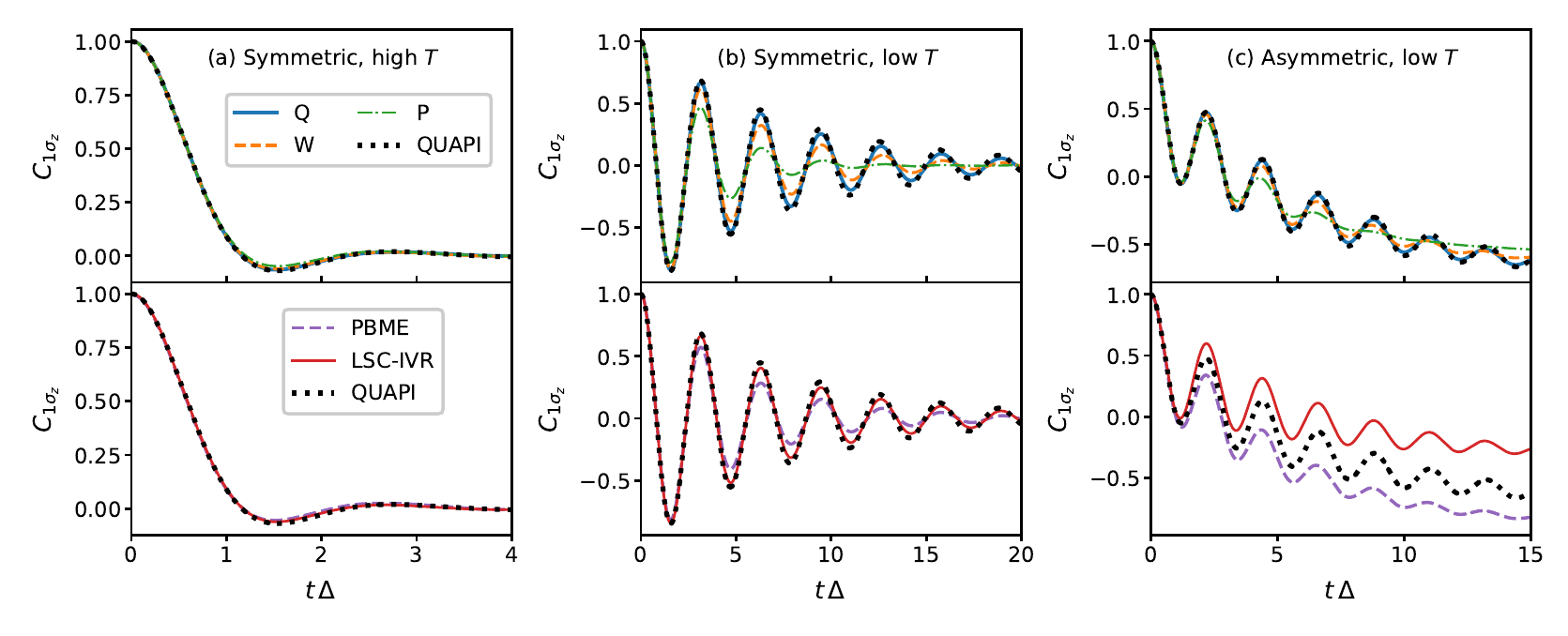}
    \includegraphics{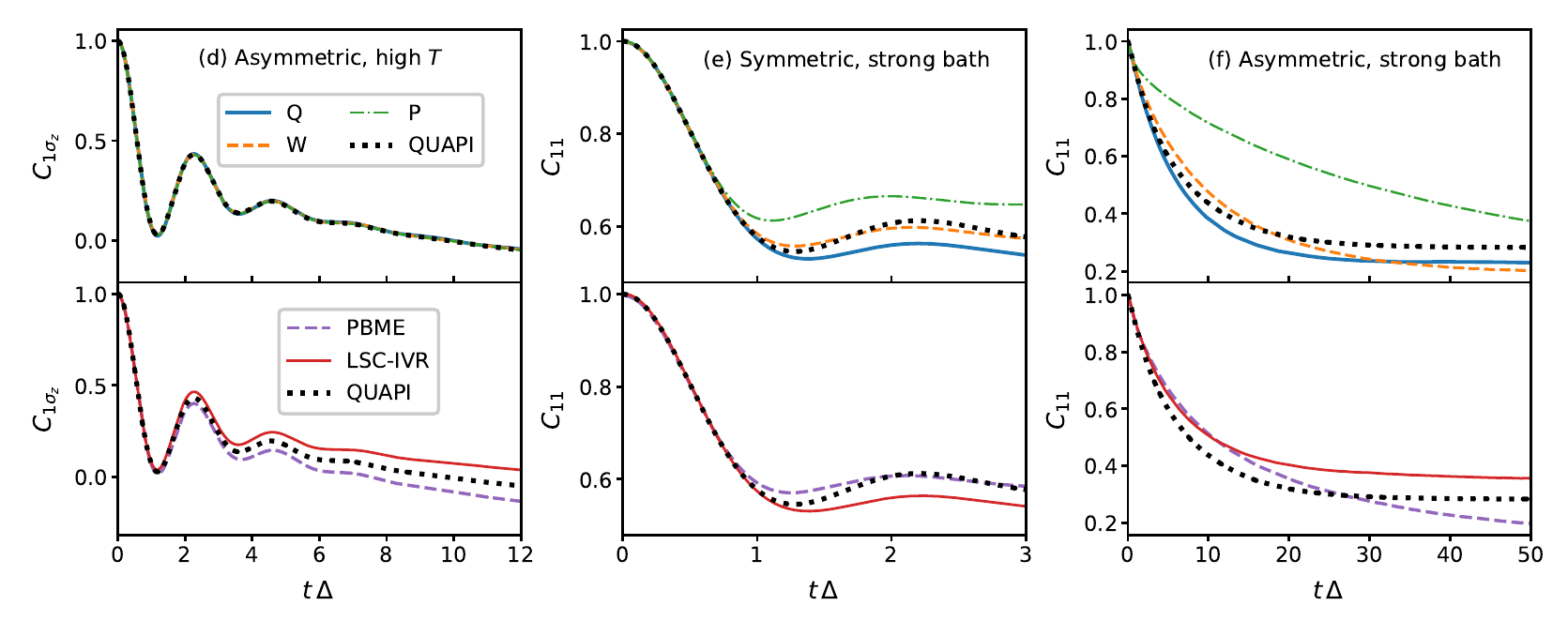}
    \caption{Time-dependent population difference $\sigma_z$ (a-d) or population of state $|1\rangle$ (e-f) after an initial projection to state $|1\rangle$, for a set of spin-boson benchmark problems. The upper panel of each pair shows the new Q-, W- and P-methods, and the lower panel shows standard MMST mapping approaches using a single (PBME) or a double (LSC-IVR) projection. Dotted lines are numerically exact path-integral results. All methods are defined in Table~\ref{tab:methods} and the model parameters are given in Table~\ref{tab:parameters}.
    }\label{fig:center}
\end{figure*}

The results for the linearized methods specified in Table~\ref{tab:methods} are compared in Figure~\ref{fig:center} with numerically exact path-integral calculations (QUAPI).\cite{makarov1994path} One sees that the Q-method captures both correct oscillations and correct long-time limits in all systems, except for the strong bath systems (e-f) where nuclear quantum effects are expected to be important. Also the W-method yields fairly accurate results with correct long-time limits, but slightly underestimates the oscillation amplitudes. This too quick decoherence is even more pronounced in the P-method. We conclude that a minimal $r_s$ in the Hamiltonian is optimal, at least for the problems considered here. The traditional MMST-based methods with double (LSC-IVR) and single (PBME) projection perform well in some cases but can be completely wrong in others, especially for final populations in asymmetric problems.

It should be pointed out that a recently published trick\cite{saller2019} to treat the identity operator in LSC-IVR or PBME gives essentially the same results as our Q-method.
There is however no direct connection between the two, since Ref.~\onlinecite{saller2019} infers no restriction on the radius $R$.
Our results suggest that the separation of the identity is what leads to correct long-time limits, whereas the choice of $\gamma$ controls the decoherences.
It should also be noted that other methods like SQC, FBTS and PLDM, which go beyond the simplest quasiclassical approximation, also perform well for the spin-boson model.\cite{Cotton2016,Hsieh2013FBTS,Huo2012PLDM} 

\begin{figure*}
    \centering
    \includegraphics{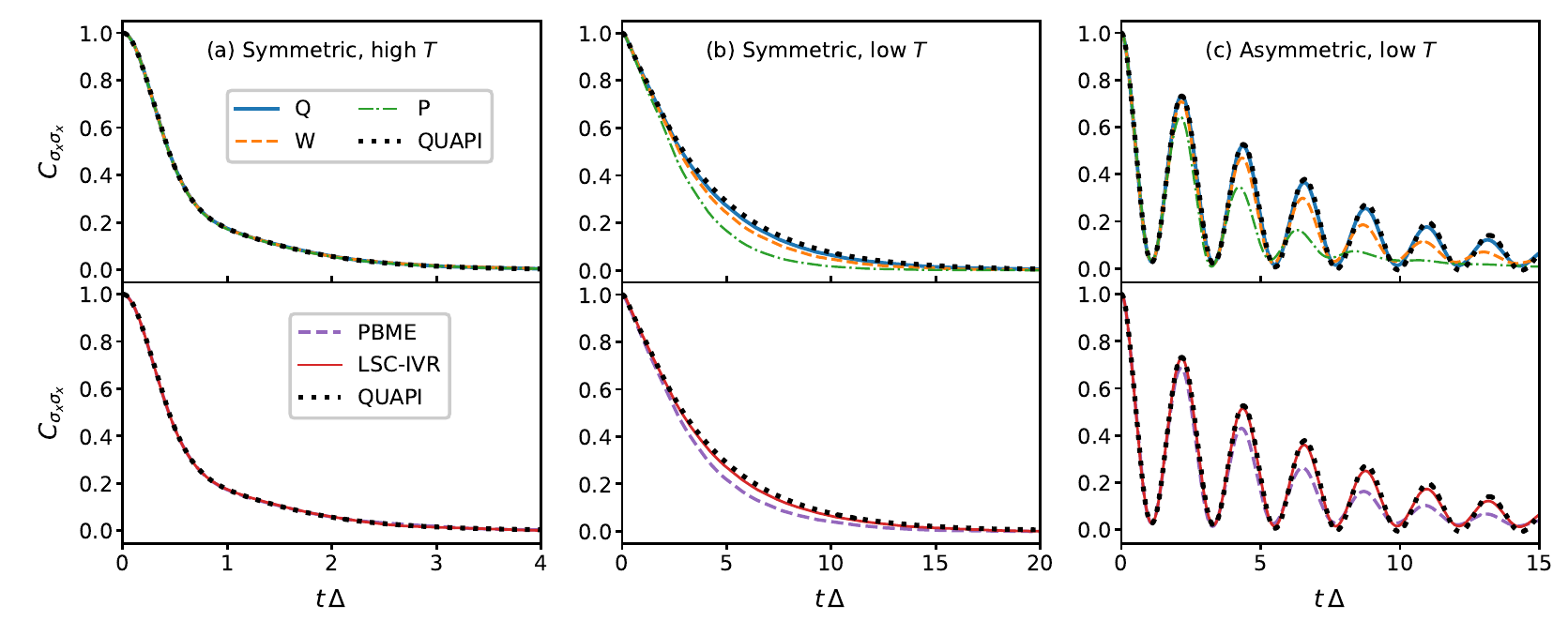}
    \includegraphics{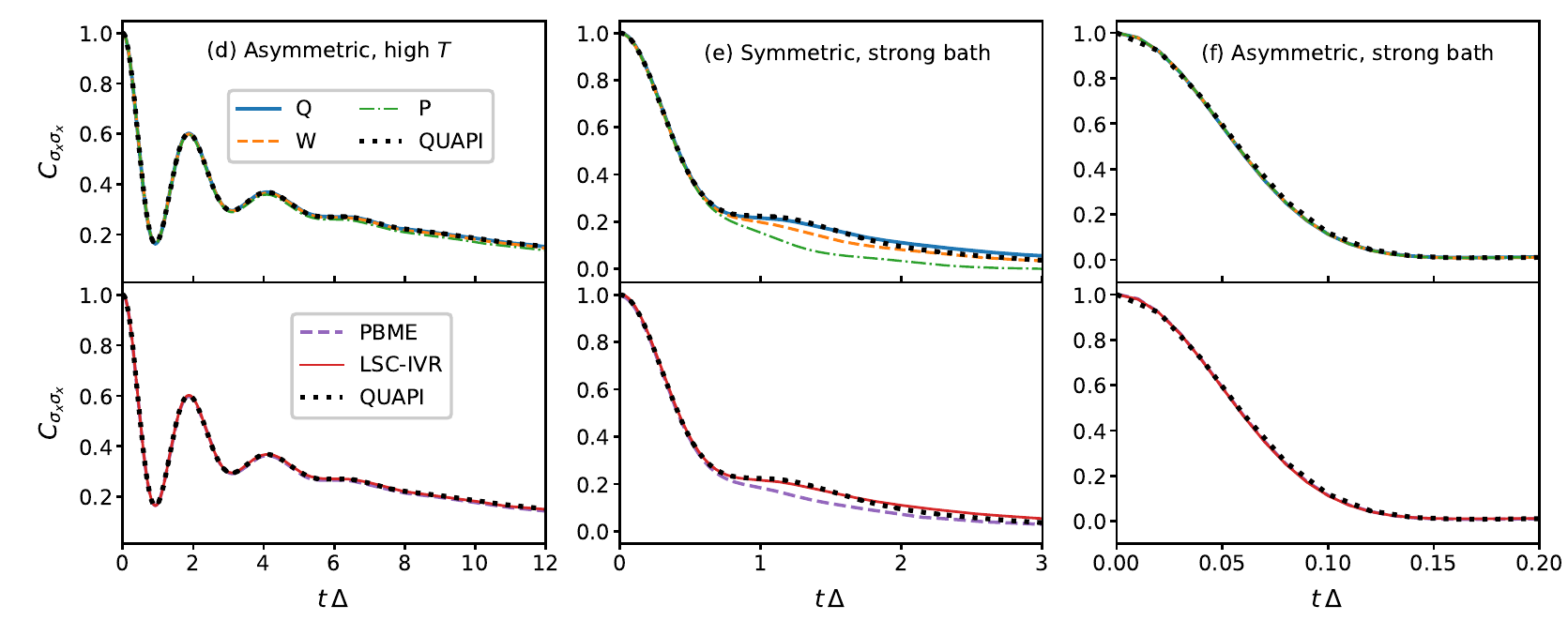}
    \caption{Same as in Figure~\ref{fig:center}, but showing the correlation function $C_{\sigma_x \sigma_x}(t)$.
    }
    \label{fig:center_xx}
\end{figure*}

When computing correlation functions of off-diagonal operators like $C_{\sigma_x\sigma_x}$ (see Fig.~\ref{fig:center_xx}), both Q and LSC-IVR are essentially exact. This is a general observation for all correlation functions of traceless observables.
Note that our method is able to compute all electronic correlation functions from the same set of trajectories, rather than different ones for each initial operator $\hat{A}$.

\begin{figure*}
    \centering
    \includegraphics{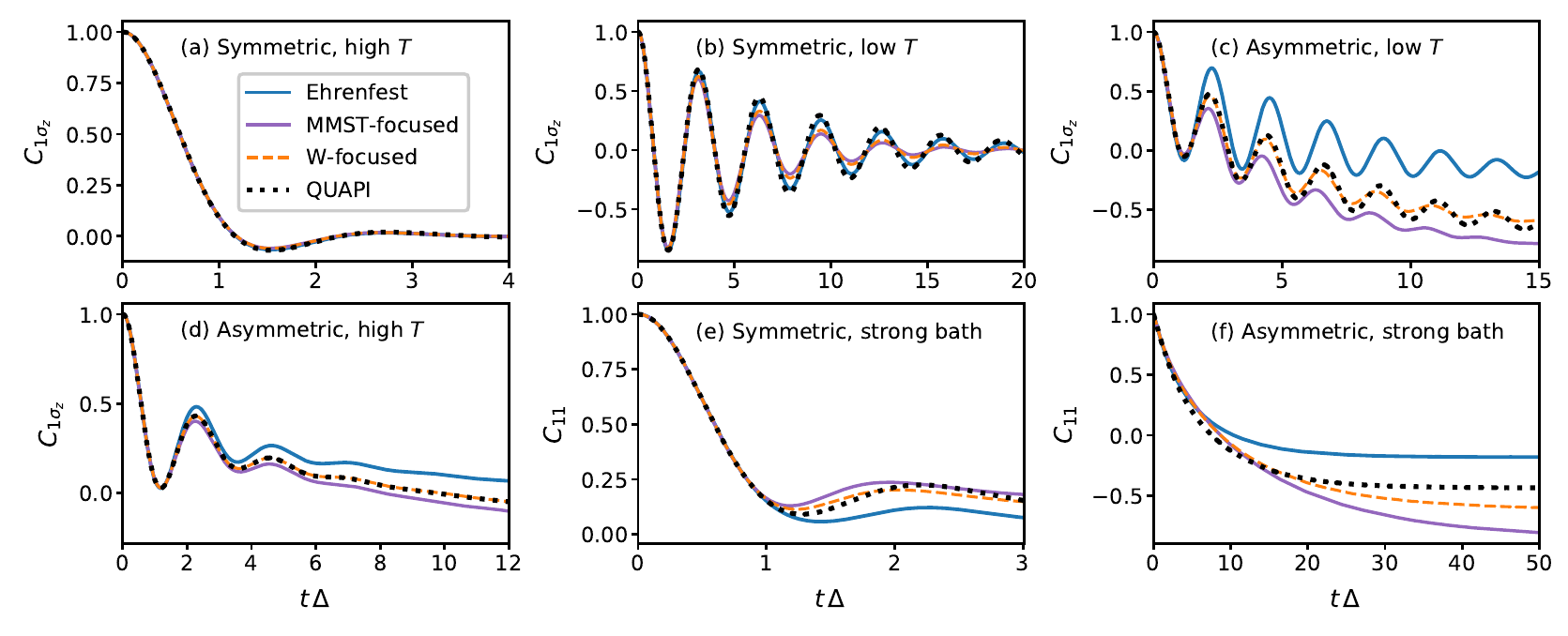}
    \caption{Same as in Figure~\ref{fig:center}, but comparing the focused methods defined in Table~\ref{tab:focused}.
    }
    \label{fig:center_foc} 
\end{figure*}

For the methods with focused initial conditions in Table~\ref{tab:focused} and Figure~\ref{fig:center_foc}, the focused W-method is clearly superior to both Ehrenfest and standard MMST ($\gamma=1)$. It uses $\gamma=\sqrt{3}-1\approx 0.732$, which is close to the values that M\"{u}ller and Stock found optimal in their simulations using action-angle (focused) initial conditions.\cite{Mueller1999pyrazine} 
Both Ehrenfest ($\gamma=0$, which is the same as a focused Q-method) and standard MMST $(\gamma=1)$ with focused initial conditions are much less accurate for asymmetric problems.
We also tried a focused P-method $(\gamma=2)$, but this deviates badly from the exact results and is not shown in the plots. This is expected from the trend seen in the asymmetric problems that a larger $\gamma$ leads to a lowering of the curve $C_{1\sigma_z}(t)$.

Our focused W-method results are of similar quality as with Cotton and Miller's symmetric quasi-classical (SQC) windowing approach, which interestingly uses the same value of $\gamma$ in their method involving square windows.\cite{Cotton2013b,Cotton2016}
Compared to SQC, however, our method does not require any choice of window functions, and no trajectories are discarded from the averaging procedure.
Also note that while the spin mapping method of Ref.~\onlinecite{Cotton2015} (which maps to two spins instead of one) has been found to give worse results than SQC defined in the MMST-basis, our spin mapping is an improvement compared to standard focused MMST. The W-function also appears to be more accurate than FBTS or PLDM with focused initial conditions.\cite{Hsieh2013FBTS,Huo2012PLDM}

We have no proof that our approach will always be better than (or as good as) these alternative methods.
However, since it avoids the problem of leaving the physical subspace, and otherwise relies on similar quasiclassical approximations, it can be expected to lead to an improvement.
The spin-mapping theory presented in this paper is limited to only two electronic levels, but since the Stratonovich-Weyl transform exists also for higher dimensional spaces, we believe it should be possible to extend to more states.
Like all quasiclassical mapping approaches, it is expected to give the wrong asymptotic limit for
trajectories emerging from a nonadiabatic crossing,
which will generally evolve on a superposition of electronic states instead of on a single state.\cite{Miller2001SCIVR}
It will also suffer from the usual classical-Wigner problem of ZPE-leakage between nuclear modes.
For studies of ultrafast dynamics this is typically an acceptable approximation.

Even though the Ehrenfest method performs poorly for these systems, its accuracy can be improved by combining it with the generalized quantum master equation.\cite{Kelly2015nonadiabatic,*Kelly2016master} In this way, observables are calculated via a memory kernel that generally decays much faster than the observables themselves, so that one can make the most out of the short-time dynamics. It would be interesting to try the same trick for our method, to compute long-time dynamics from short simulations and (potentially) further improve the accuracy.

All together, we find that the most accurate method for these spin-boson systems is the Q-method without focusing. This only fails to quantitatively reproduce the exact result for the strong bath, where nuclear quantum effects are expected to be important.
Further tests will be carried out to discover if this behaviour carries over to other systems.


\section{Conclusions and future prospect}
We have presented a family of three new mappings for two-level molecules, based on the Stratonovich-Weyl transform of spin-\thalf\ systems. Just like the MMST mapping, these are exact on a quantum-mechanical level. In a quasiclassical treatment, both the spin and MMST mappings give exact results for a bare two-level system, but the two approaches have different accuracy when there is a coupling to an environment.

By using the traciality property of the Stratonovich-Weyl transform [Eqs.~\eqref{eq:trAB} and~\eqref{eq:trAB_W}] we can approximate electronic correlation functions in a manner similar to classical Wigner approaches. The Q-method seems particularly promising because it gives accurate results and avoids the possibility of inverted potentials.
The self-dual W-representation also appears to be very useful, in particular in implementations using focused initial conditions.



Unlike most other quasiclassical methods, our scheme only requires two free real variables to represent the two-level system, so that it has the same number of (electronic) degrees of freedom as the quantum problem. The Hamiltonian is unique, without ambiguity in the splitting between the state-dependent and state-independent potentials. Via a coordinate transformation we show that the same dynamics can be generated with the usual quadratic MMST Hamiltonian, leading to numerically stable linear equations of motion. 
Our approach is therefore closely related to the MMST mapping but with a couple of important differences. 
Firstly, the total population is always fixed to one, because the $\{\mathsf{X},\mathsf{P}\}$ variables are constrained to a three-dimensional sphere.
Secondly, it uses a different value of the ZPE-parameter, $\gamma$, which scales the operators in a way that is unexpected from an MMST point of view. Because of this, it is not enough to simply fix the total population in an MMST simulation $(\gamma=1)$, since one would still need to rescale the traceless part of the $\hat{B}$ operator by $3/4$. 

A fundamental advantage compared to MMST is that the system cannot leave the physical subspace, so there is no need for additional projections.
We believe that this fact will be particularly useful for the development of a nonadiabatic version of ring-polymer molecular dynamics (RPMD),\cite{Ananth2013MVRPMD,Chowdhury2017coherent,mapping,richardson2017vibronic} which would add nuclear quantum effects to the simulations, as well as to allow sampling from an exact thermal distribution.


\begin{acknowledgments}
We are grateful for Michael Thoss for valuable comments on the manuscript.
J.E.R.\ is supported by the Hans H. G\"{u}nthard scholarship, 
and both authors
acknowledge support from the Swiss National Science Foundation through the NCCR MUST (Molecular Ultrafast Science and Technology) Network.
\end{acknowledgments}

\appendix

\section{Derivation of the equations of motion from the QCLE}\label{sec:app_QCLE}
The quantum-classical Liouville equation\cite{kapral1999mixed} is an equation of motion for mixed quantum-classical dynamics that, in its full implementation, is exact for spin-boson models.\cite{kapral2015quantum} It has previously been used to analyse both surface hopping methods\cite{kapral1999mixed,horenko2002,kelly2013efficient} and methods based on the MMST representation.\cite{Kim2008Liouville,nassimi2010,Kelly2012mapping} After performing a partial Wigner transform of the nuclear coordinates, it states that the time evolution of an operator $\hat{B}=\hat{B}(x,p)$ follows the equation
\begin{equation}\label{eq:qcle}
    \frac{\rd}{\rd t}\hat{B} = \ii \hat{\mathcal{L}} \hat{B} \equiv \ii[\hat{H},\hat{B}]-\tfrac{1}{2}(\{\hat{H},\hat{B}\}-\{\hat{B},\hat{H}\}),
\end{equation}
where $[\cdot,\cdot]$ indicates the commutator and $\{\hat{A},\hat{B}\}=\pd{\hat{A}}{x}\pd{\hat{B}}{p}-\pd{\hat{A}}{p}\pd{\hat{B}}{x}$ is the Poisson bracket in the nuclear phase space. We shall transform this to the Stratonovich-Weyl representation by using its definition
\begin{equation}\label{eq:SW}
    B_s(\bm{u}) = \tr[\hat{B}\hat{w}_s(\bm{u})] = B_0 + r_s \bm{B}\cdot \bm{u},
\end{equation}
for a general decomposition $\hat{B}=B_0\hat{\mathcal{I}}+\tfrac{1}{2}\bm{B}\cdot\hat{\bm{\sigma}}$,
together with the inversion formula
\begin{equation}\label{eq:SWinverse}
    \hat{B} = \int \rd\bm{u} \, \hat{w}_{{\bar{s}}}(\bm{u}) B_{s}(\bm{u}).
\end{equation}
The left-hand side of Eq.~\eqref{eq:qcle} is directly written as
\begin{equation} \label{eq:lhs}
    \frac{\rd}{\rd t}\hat{B} = \int \rd\bm{u}\, \hat{w}_{{\bar{s}}}(\bm{u}) \frac{\rd}{\rd t} B_s(\bm{u}).
\end{equation}
For the right-hand side we first need the SW-transform of a product.
Repeated use of Eqs.~\eqref{eq:SW}--\eqref{eq:SWinverse} gives
\begin{multline}
    [\hat{A}\hat{B}]_s(\bm{u}) = \int \rd\bm{u}' \,\rd\bm{u}''\, \tr[\hat{w}_s(\bm{u})\hat{w}_s(\bm{u}')\hat{w}_s(\bm{u}'')] \\ \times A_{{\bar{s}}}(\bm{u}')B_{{\bar{s}}}(\bm{u}'').
\end{multline}
Using the result
\begin{multline}
    \tr[\hat{w}_s(\bm{u})\hat{w}_s(\bm{u}')\hat{w}_s(\bm{u}'')] = \tfrac{1}{4}+r_s^2(\bm{u}\cdot\bm{u}' + \bm{u}\cdot\bm{u}'' + \bm{u}'\cdot\bm{u}'') \\ 
    + 2\ii r_s^3\bm{u}\cdot(\bm{u}'\times\bm{u}'')
\end{multline}
together with $r_s r_{{\bar{s}}}=3/4$, we can (after some algebra) express the commutator as
\begin{equation}\label{eq:rhs1}
\ii [\hat{H},\hat{B}] = r_s \int  \rd\bm{u} \, \hat{w}_{\bar{s}}(\bm{u}) \, \bm{B} \cdot (\bm{H}\times \bm{u}),
\end{equation}
where we used the integrals
\begin{equation}
    \int \rd\bm{u} = 2, \quad \int \rd\bm{u}\, u_i = 0, \quad \int \rd\bm{u}\,u_iu_j = \tfrac{2}{3}\delta_{ij}.
\end{equation}
Similarly for the Poisson bracket we get
\begin{multline}\label{eq:rhs2}
     -\tfrac{1}{2}(\{\hat{H},\hat{B}\}-\{\hat{B},\hat{H}\}) = \int \rd\bm{u}\,\hat{w}_{\bar{s}}(\bm{u})\times \\
     \left[\left(\frac{p}{m}\pd{}{x} - \pd{H_0}{x} \pd{}{p} - r_s\pd{\bm{H}}{x}\cdot\bm{u}\pd{}{p} \right)B_s(\bm{u})   \right. \\
     +\left. \left( r_s^2\pd{\bm{H}}{x}\cdot\bm{u}\pd{\bm{B}}{p}\cdot\bm{u}-\frac{1}{4}\pd{\bm{H}}{x}\cdot\pd{\bm{B}}{p} \right)\right].
\end{multline}
We will omit the terms on the last line in order to find a solution of uncoupled trajectories. Such an approximation is used also to derive PBME,\cite{Kelly2012mapping} but compared to the MMST-basis used there, the Stratonovich-Weyl representation has the advantage that it requires no projectors. It is therefore not possible for the system to leave the physical subspace, even without the last line in Eq.~\eqref{eq:rhs2}. A detailed analysis shows that the omitted term integrates to
\begin{equation}
    \hat{\mathcal{I}}\left(\frac{r_s^2}{3}-\frac{1}{4}\right)\pd{\bm{H}}{x}\cdot\pd{\bm{B}}{p}
\end{equation}
which interestingly vanishes for $r_s=\sqrt{3}/2=r_\text{W}$, indicating a potential advantage of the symmetric W-function. It is nevertheless an approximation to omit this term in an uncoupled trajectory-solution, since it does not vanish for all higher derivatives.

Combining all the terms of Eq.~\eqref{eq:qcle} together, we get
\begin{multline}\label{eq:QCLEend}
    \int \rd\bm{u}\,\hat{w}_{\bar{s}}(\bm{u})\frac{\rd}{\rd t}B_s(\bm{u}) = \int \rd\bm{u}\,\hat{w}_{\bar{s}}(\bm{u}) \bigg[ r_s\bm{B}\cdot(\bm{H}\times\bm{u}) +  \\ +\left.\left(\frac{p}{m}\pd{}{x} - \pd{H_0}{x}\pd{}{p} - r_s\pd{\bm{H}}{x}\cdot\bm{u}\pd{}{p} \right)B_s(\bm{u})\right].
\end{multline}
The easiest solution to this equation is obtained by setting the integrands on each side to be equal.
Using the chain rule to expand the total derivative of $B_s(\bm{u},x,p)$ as
\begin{equation}
    \frac{\rd}{\rd t}B_s = \pd{B_s}{\bm{u}}\cdot\dot{\bm{u}}+\pd{B_s}{x}\dot{x}+\pd{B_s}{p}\dot{p},
\end{equation}
and comparing the expressions term by term,
we obtain the equations of motion~\eqref{eq:eom_all}.
Therefore an ensemble of these uncoupled trajectories
can be used to obtain quasiclassical dynamics as described in the main text.

\end{document}